# An ion-independent phenomenological relative biological effectiveness (RBE) model for proton therapy

Liheng Tian[1], Christian Hahn[1,2,3], Armin Lühr[1]

[1]TU Dortmund University, Department of Physics, Dortmund, Germany

[2]OncoRay, National Center for Radiation Research in Oncology, Faculty of Medicine

and University Hospital Carl Gustav Carus, Technische Universität Dresden,

Helmholtz-Zentrum Dresden-Rossendorf, Dresden, Germany

[3] Department of Radiotherapy and Radiation Oncology, Faculty of Medicine and University Hospital Carl

Gustav Carus, Technische Universität Dresden, Dresden, Germany

E-mail: Liheng.Tian@tu-dortmund.de

**Abstract**:

Background: A relative biological effectiveness (RBE) of 1.1 is used for proton therapy though clinical evidence of varying RBE was raised. Clinical studies on RBE variability have been conducted for decades for carbon radiation, which could advance the understanding of the clinical proton RBE given an ion-independent RBE model. In this work, such a model, linear and simple, using the beam quantity $Q = Z^2/E$ ($Z$ = ion charge, $E$ = kinetic energy per nucleon) was tested and compared to the commonly used, proton-specific and linear energy transfer (LET) based Wedenberg RBE model.

Material and methods: The Wedenberg and Q models, both predicting $RBE_{max}$ and $RBE_{min}$ (i.e., RBE at vanishing and very high dose, respectively), are compared in terms of ion-dependence and prediction power. An experimental in-vitro data ensemble covering 115 publications for various ions was used as dataset.

Results: The model parameter of the Q model was observed to be similar for different ions (in contrast to LET). The Q model was trained without any prior knowledge of proton data. For proton RBE, the differences between experimental data and corresponding predictions of the Wedenberg or the Q model were highly comparable.

Conclusions: A simple linear RBE model using Q instead of LET was proposed and tested to be able to predict proton RBE using model parameter trained based on only RBE data of other particles in a clinical proton energy range for a large in-vitro dataset. Adding (pre)clinical knowledge from carbon ion therapy may, therefore, reduce the dominating biological uncertainty in proton RBE modelling. This would translate in reduced RBE related uncertainty in proton therapy treatment planning.

# 1. Introduction

A fixed relative biological effectiveness (RBE) of 1.1 is routinely used in clinical proton therapy [1]. Clinical research reported that the toxicity at the end of the spread-out Bragg-peak (SOBP) was related to an increased RBE [2,3]. The RBE is believed to be affected by e.g. linear energy transfer (LET) and tissue sensitivity [4–6]. The applied assumption of a constant RBE might be reasonable if an average for an SOBP field is assigned, as the average cell survival RBE was found to be around 1.1 in the entrance and 1.15 in the center [5]. However, at the end of an SOBP, the average RBE increases substantially with RBE around 1.35 at the distal edge and 1.7 in the distal-falloff [5] and some published studies estimated even higher values of, e.g., 2-3 [7]. Additionally, emerging clinical evidences pointed out that the difference between the actual RBE and the applied constant value is likely to be of clinical relevance [2–5,8–11].

Different RBE models have been proposed. Phenomenological models rely on the fitting of available experimental measurements [12–17]. Thus, their clinical application is affected by the limited amount of clinical proton data. Mechanistic models are based on generally believed mechanisms, namely, that the enhanced RBE of ion irradiation is caused by the microscopic dose distribution in the cell [18–20]. However, these models are mainly applied in heavy-ion radiotherapy.

In clinical carbon ion therapy, the RBE plays a central role and has been researched for decades [21–24]. Its importance stems from the fact that carbon RBE, and especially its variability, is much more pronounced in clinical SOBP (RBE between 2 and 5) than for protons, due to the six times higher atomic charge of carbon ions. Therefore, learning directly from clinical carbon RBE data and using this knowledge to advance RBE research and treatment for proton therapy appears highly attractive. To achieve this goal, an (preferably simple) ion-independent RBE description is desirable. Recently, the use of the beam quality Q [25,26], as an alternative to the conventional LET, was proposed and hypothesized to lead to an ion-independent RBE model [26] as it relates to the radial dose distribution around the ion track rather than its integral energy loss per step given by the LET. Beam quality was defined as $Q = Z^2/E$, with Z and E being the ion's charge and kinetic energy per nucleon, respectively.

In this work, an RBE model based on Q was established and compared to a conventional LET model. The particle irradiation data ensemble (PIDE), recording 115 publications of different experimental in-vitro RBE data for various ions [27] [28], was used to build and test the Q model in terms of ion dependency and prediction power for proton RBE.

## 2. Material and methods

### 2.1 PIDE dataset and data selection

PIDE (version: 3.2) [28] records the experimental data of in-vitro cell survival experiments of 115 publications covering 1118 data points of 21 types of ion irradiation.

In this work, the maximum and minimum RBE values given by $RBE_{max}=\alpha_i/\alpha_x$, $RBE_{min}=\sqrt{\beta_i/\beta_x}$, respectively [12,29], are modelled using clinically accessible quantities that are available in PIDE, i.e. the biological quantity $\alpha_x/\beta_x$ and either LET or Q as physical quantity. Here, $\alpha_i$, $\alpha_x$, $\beta_i$ and $\beta_x$ are the α and β values of ion and photon irradiation, respectively, according to the linear quadratic (LQ) model. To filter data from PIDE, firstly, proton data with $RBE_{max} > 35$ were excluded (cell line PDVC57 [30]), as they are significantly higher than all other recorded proton data (<9). The remaining data were subject to a selection workflow (figure 1) and considered for analysis if fulfilling the following conditions: 1) positive and finite $\alpha_x/\beta_x$, 2) asynchronous cell cycle distribution, 3) monoenergetic irradiation, 4) $Q < 2.5$ $(A \cdot MeV)^{-1}$ and LET lower than 30, 103, 208 and 233 keV/μm for proton, helium, carbon and neon, respectively, avoiding potential 'over-killing' [26,31]. 5) $Q/(\alpha_x/\beta_x)$ and $LET/(\alpha_x/\beta_x)$ (explanation cf. section 2.2 and 2.3) values within the interval covered by proton data for the corresponding model; 6) at least five data points for the same ion type. While for protons the same dataset was considered for LET and Q modelling, for other ions, the resulting datasets could slightly differ due to condition 5) (cf. figure 1 and supplement figure S1).

LET values were directly taken from PIDE irrespective of LET definition (i.e., regardless of dose/track averaged LET). Other quantities such as Q, $Q/(\alpha_x/\beta_x)$ and $LET/(\alpha_x/\beta_x)$ were calculated from the tabulated PIDE records. Note, not all underlying publications provided, both, E and LET values. The missing values were calculated by the PIDE group based on the reported counterpart value using the software ATIMA [25]. In PIDE, two types of α and β are reported: first, the originally published data and, second, data retrospectively obtained by the PIDE group by LQ fitting of the underlying radiation response data. Here, publication-reported α and β values were used to ensure consistency with the original Wedenberg model [16].

## 2.2 Correlation analysis

The Spearman's correlation coefficient, ρ, was calculated (Pandas package [32]) between different clinically available input (Q, LET, $\alpha_x/\beta_x$, $\alpha_x$ and $\beta_x$) and output ($RBE_{max}$, $RBE_{min}$, $\alpha_i$, and $\beta_i$) quantities. $Q/(\alpha_x/\beta_x)$ and its LET counterpart of $LET/(\alpha_x/\beta_x)$ were also considered as input quantities. Other quantities of $LET^n \cdot (\alpha_x/\beta_x)^m$ with *m* and *n* within [-3,3] (e.g., $LET^3/(\alpha_x/\beta_x)$ for *n*=3 and *m*=-1, i.e., Peeler model [33]) showed lower ρ values and, thus, are not further discussed in this manuscript.

## 2.3 Q model and LET model

Due to the observed high correlation ρ between $Q/(\alpha_x/\beta_x)$ and $RBE_{max}$ and low ρ between $RBE_{min}$ and all available inputs (cf. figure 2), the Q model was proposed as:

$$RBE_{max} = 1 + k_Q \cdot \frac{Q}{\alpha_x/\beta_x} \qquad (1)$$

$$RBE_{min} = 1 \qquad (2)$$

Correspondingly, high ρ between $LET/(\alpha_x/\beta_x)$ and $RBE_{max}$ suggested the following LET-driven model for comparison:

$$RBE_{max} = 1 + k_{LET} \cdot \frac{LET}{\alpha_x/\beta_x} \qquad (3)$$

$$RBE_{min} = 1 \qquad (4)$$

$k_Q$ and $k_{LET}$ are the model parameters and $Q/(\alpha_x/\beta_x)$ and $LET/(\alpha_x/\beta_x)$ the inputs of the Q and LET model, respectively. Note, the form of the LET-driven model corresponds to the Wedenberg model [16] but the model parameter $k_{LET}$ may differ.

Expressions for $\alpha_i$ and $\beta_i$ follow directly from the definition of $RBE_{max}$, $RBE_{min}$ and the proposed Q model:

$$\alpha_i = \alpha_x \cdot RBE_{max} = \alpha_x + k_Q \cdot \beta_x \cdot Q \quad \text{and} \quad \beta_i = \beta_x \cdot RBE_{min}^2 = \beta_x \qquad (5)$$

For a given photon or proton dose $D_x$ or $D_i$, respectively, RBE can be expressed by the Q model as (cf. supplement):

$$RBE(Q, D_x, \alpha_x/\beta_x) = \frac{\alpha_x/\beta_x + k_Q \cdot Q + \sqrt{(\alpha_x/\beta_x + k_Q \cdot Q)^2 + 4D_x \cdot (\alpha_x/\beta_x + D_x)}}{2(\alpha_x/\beta_x + D_x)} \qquad (6)$$

and

$$RBE(Q, D_i, \alpha_x/\beta_x) = -\frac{R_x}{2D_i} + \frac{1}{D_i} \cdot \sqrt{\frac{1}{4}(\alpha_x/\beta_x)^2 + (\alpha_x/\beta_x + k_Q \cdot Q) \cdot D_i + D_i^2} \qquad (7)$$

## 2.4 RBE trend of different ions

The $RBE_{max}$ trends of different ions were investigated by fitting both the Q model, Eq. (1), and the LET model, Eq. (3) using the datasets of individual ions as well as combined datasets pooling data from all particle types (cf. figure 1 and supplement figure S1 for different datasets used for each analysis). The slopes $k_Q$ and $k_{LET}$ and corresponding $r^2$ values of the linear regressions were compared for different ions to determine the ion dependency of the models. Additionally, the residuals between the experimental data and the global fitting (fitting using all data regardless of particle type) were determined. The ANOVA (analysis of variance) test was applied to investigate if the residuals of different particle types were distributed differently.

To test the assumption $RBE_{min} = 1$ (Eqs. (2) and (4)) for individual ions, the trend of change in $RBE_{min}$ with $LET/(\alpha_x/\beta_x)$ and $Q/(\alpha_x/\beta_x)$ was investigated by linear regression. The mean values for $RBE_{min}$ were calculated. PIDE data with missing $\beta_i$ were excluded from this analysis (cf. figure 1).

## 2.5 Model prediction

For the Q model, first, the dataset was split into a training set (of 137 data points including 34 helium, 88 carbon and 15 neon ions) and a test set (of 48 proton data points, Supplement figure S1). Second, the model parameter $k_Q$ was obtained by fitting the model (Eq. (1)) to the training set. Using the $k_Q$ determined from all ions but protons, the proton $RBE_{max,Q}$ values were predicted for the $Q/(\alpha_x/\beta_x)$ values of all considered experimental proton data. $\alpha_i$ values were predicted according to Eq. (5) using the same $k_Q$ parameter value. The competing LET model prediction values, $RBE_{max,LET}$ were calculated with the original Wedenberg model ($k_{LET}$=0.434±0.068 Gy·(keV/μm)$^{-1}$) [16] for the same proton data as test set.

The prediction power of the RBE models was measured by the $r^2$ between the prediction and the experimental records for both $RBE_{max}$ and $\alpha_i$ prediction.

## 3. Results

Following the proposed data selection criteria, four types of ion irradiation were eligible: proton, helium, carbon and neon (figure 1) with maximum LET values of 27.6, 74.6, 117.0, 116.0 keV/μm, respectively. Data of other ions had less than 5 data points and, thus, were not further considered. For the Q model fitting, 185 data points with $\alpha_x/\beta_x$ within [0.66 Gy, 69.5 Gy] were used for $RBE_{max}$ and 115 data points were used for $RBE_{min}$. For the LET model fitting, 164 data points with $\alpha_x/\beta_x$ within [1.4 Gy, 69.4 Gy] were used for $RBE_{max}$ and 105 data points were available for $RBE_{min}$. Data set sizes as well as histograms of $\alpha_x/\beta_x$, E, LET, Q, LET/$(\alpha_x/\beta_x)$, and $Q/(\alpha_x/\beta_x)$ distributions are shown in Supplement figures S1 and S2.

Spearman's correlation between $RBE_{max}$ and $Q/(\alpha_x/\beta_x)$ ($\rho$=0.82) was observed to be higher than for all others quantities including LET/$(\alpha_x/\beta_x)$ (0.76) cf. figure 2. Correlation between $RBE_{min}$ and any Q or LET related input variable was low ($|\rho|\leq0.34$).

$RBE_{max}$ was observed to increase linearly with $Q/(\alpha_x/\beta_x)$ (figure 3) and $k_Q$ was similar for different ions. The linear slopes for individual particles were tested to be not significantly different from the global slope (p>0.55, supplement table S1). Thus, for Eq.(1) a common $k_Q$ = (15.5±0.5) A·MeV·Gy could be applied to the total Q model dataset (pooling all particles). For each considered particle type separately, $RBE_{max}$ was also observed to increase linearly with LET/$(\alpha_x/\beta_x)$. But, in contrast to the Q model, the $k_{LET}$ was ion-dependent (p ≤ 0.005) and a global $k_{LET}$ was not applicable. This was confirmed by the ANOVA test: the residuals between the experimental data and the global fitting of the Q model were not significantly ion-dependent (p=0.63), while they were significantly ion-dependent for the LET model (p=3.4×10$^{-7}$).

In line with the low correlation between RBE$_{min}$ and all considered input parameters (figure 2), no clear trend of the change of RBE$_{min}$ following the change of the $Q/(\alpha_x/\beta_x)$ or $LET/(\alpha_x/\beta_x)$ was observed (r$^2$≤0.05, figure 4). For each of the individual particles, neither the slopes of the regressions between RBE$_{min}$ and $Q/(\alpha_x/\beta_x)$ or $LET/(\alpha_x/\beta_x)$ were significantly different from 0, nor the mean value of RBE$_{min}$ was significantly different from 1.

The Q model predicted the experimental proton RBE$_{max}$ data with a similar error level (r$^2$=0.72) as that of the widely-used proton specific Wedenberg model (r$^2$=0.73, figure 5). Note that the model parameter k$_Q$ of the predictive Q model discussed here was exclusively trained on RBE knowledge of other particles without any input of proton RBE data. Differences between the predictions by the Q model (RBE$_{max,Q}$) or the Wedenberg model (RBE$_{max,LET}$) and the experimental records were comparable (supplement figure S3) and distributed with mean (standard deviation, SD) values of -0.07 (±0.94) and -0.21 (±0.91) for RBE$_{max,Q}$ and RBE$_{max,LET}$, respectively. Systematic errors were observed for experiments with RBE$_{max}$<1 as both models cannot take values below 1 due to the employed model formula.

Using the same Q model, the predictions of proton $\alpha_i$ values were observed to match the experimental records (figure 6). The differences between predicted and experimentally reported $\alpha_i$ were distributed with a mean (SD) value of 0.01 (±0.12) Gy$^{-1}$ and r$^2$ of 0.79 (figure 6 B), when only considering $\alpha_i$ < 1 Gy$^{-1}$. For the same $\alpha_i$ range, the Wedenberg model resulted in corresponding mean (SD) and r$^2$ values of 0.07 (±0.12) Gy$^{-1}$ and 0.76, respectively.

## 4. Discussion

This work tested the ability of the Q model to describe and predict RBE for different light ions based on the extensive in-vitro dataset PIDE and compared it to an established LET model. RBE$_{max}$ was found to be highly correlated to the quantity $Q/(\alpha_x/\beta_x)$ regardless of particle type. The Q model for RBE$_{max}$, which is linear in $Q/(\alpha_x/\beta_x)$, was tested to be not significantly ion-dependent. The Q model trained without any RBE data of proton, i.e., using only data of other particles, was able to make predictions for protons with similar precision as the widely used Wedenberg RBE model [16], which, in contrast, was directly built on proton data. Exploiting the potential of Q, the amount of radiobiological data available for RBE modelling increases substantially when compared with conventional ion-dependent LET modelling approaches, which are restricted to data from single ions. This may be especially relevant for clinically used proton or helium irradiation, i.e., if ion-independence will further be verified in a clinical setting, the RBE knowledge from carbon irradiation could be transferred to proton and helium ions. Additionally, an enlarged RBE dataset could be divided into training and testing datasets allowing the Q model to be evaluated in terms of prediction power. In contrast, most of the empirical LET-based proton RBE models, including the Wedenberg model, resulted from a direct fit of the available proton RBE data, which is actually a subset of the proton data in PIDE.

Based on the analyzed PIDE data, a Q model given by Eqs (3) and (4) with a $k_Q$ value of (15.5±0.5) A·MeV·Gy is proposed. For the LET model and protons, the $k_{LET}$ reported by Wedenberg et al. was (0.434±0.068) Gy·(keV/μm)$^{-1}$ [16]. Here, an updated $k_{LET}$=(0.49±0.03) Gy·(keV/μm)$^{-1}$ was obtained using the same methods but an extended proton dataset. Although these two values are not significantly different, it is suggested to prefer the updated $k_{LET}$ when applying the Wedenberg approach in the future, due to the enlarged dataset.

The correlation between RBE$_{min}$ or $\beta_i$ and all considered input quantities was found to be low (figure 2 and 4). Available RBE models have been controversial in this point proposing different behaviors for $\beta_i$, including an increase, a decrease or no change with increasing LET [27,34]. Recently, it had been reported that experimental $\beta_i$ values were affected by the experimental design and strategy of LQ model fitting [35], which may add further uncertainty and complicates analysis of experimental data. While more research is needed to improve modeling of $\beta_i$ [27], here, no significant deviation from a constant RBE$_{min}$ value of 1 or between particle types could be demonstrated. Thus, the assumption of $\beta_i = \beta_x$ was maintained.

For proton irradiation, thirteen different phenomenological variable RBE models are compared in [15], showing agreement in the LET dependence of proton RBE but disagreement in the role of $\alpha_x/\beta_x$.

In this work, RBE$_{max}$ was observed to increase with LET and decrease with $\alpha_x/\beta_x$, which corresponded to the models by Carabe [12], Mairani [13], McNamara [14], Wedenberg [16], Tilly [17] and Rørvik [34], which are all proton-specific. In contrast, the Q model does not rely on the limited proton RBE data alone. Compared to the LET, Q is easy to calculate as it is independent of any material properties [36] and requires only the charge and energy of an ion, which are available within treatment planning systems. Though LET and Q are highly correlated for a single particle type and given material, for mixed particle spectra, which are always present in clinical treatment fields, the observed similarity of model parameters ($k_Q$) for different ions may allow for a useful voxel-wise averaging of Q for different particles using the same weight. In contrast, LET averaging in a voxel is complicated by the missing ion-independence of the LET-RBE relation. Finally, the beam quality Q only characterizes the ion irradiation while LET also strongly depends on the properties of the considered medium. Other RBE models, e.g. the Local Effect Model [20] and the Microdosimetric-Kinetic Model [19] are mechanistic and clinically applied for carbon ion therapy. These models demonstrate that the enhancement of the RBE of ion irradiation is caused by the microscopic local dose distribution near the track of the incident particle in a cell. Q is believed to be related to the radial dose around an ion track [25,26].

While radiobiological data can be associated with substantial uncertainties and experimental settings differ among institutes, no experimental uncertainties are recorded by the PIDE database. Similarly, various LET definitions were used by different RBE experiments but not always fully specified [36] and energies of some data points were only calculated by the PIDE group. Inconsistencies between experimental literature data may be partially overcome by large datasets [28]. Here, the proposed Q model, which used in total 185 data points, may provide a substantial advantage over other empirical LET models, which are restricted to proton data only (here 48 points).

In this work, the Q model was only studied within the $Q/(\alpha_x/\beta_x)$ interval (cf. supplement figure S2) that is spanned by the considered proton data in a clinically relevant range [11]. The considered proton energy ranged from 1 to 164 MeV (0.5 - 26 keV/µm), while high-LET irradiation as relevant for carbon ion therapy was not involved. It is well known that an overkill sets in at high LET values. Therefore, in regions with higher $Q/(\alpha_x/\beta_x)$, it is expected that the observed linearity between RBE and

Q decreases, uncertainty increases, and a nonlinear model becomes necessary [27,28]. Here, the lower energy threshold for protons was set to 1 MeV [$Q = 1$ (A·MeV)$^{-1}$] corresponding to a maximum proton range of less than 25 µm in water [37], i.e., in the order of the size of a single cell. Cell irradiation experiments with shorter ranges become increasingly difficult and may be affected by high experimental uncertainties. Using Q, end of range proton RBE data may, instead, be estimated based on data from heavier ions, which have a much longer range at the same Q. Such data for protons may particularly be interesting for radiation-induced toxicity.

For the Q model, most of the $\alpha_i$ predictions were within an error of 0.25 Gy$^{-1}$ except for data of two single cell lines AG01522 [7] and HCT116 [38]. The reported $\alpha_x$ value for HCT116 (1.39 Gy$^{-1}$) was an outlier being much higher than $\alpha_x$ from all other experiments (all <0.75 Gy$^{-1}$; mean=0.24 (±0.20) Gy$^{-1}$). The experimental $\alpha_i$ records for AG01522 (>1 Gy$^{-1}$) were the highest compared to all other records (<0.9 Gy$^{-1}$; mean=0.41 (±0.24) Gy$^{-1}$).

This study considered only mono-energetic irradiation. Future work, therefore, needs to focus on considering Q in an SOBP to quantify the impact of clinically relevant energy and secondary particle spectra and on in vivo and clinical data. Previously, the Q model had successfully been applied [26] to model RBE of fractionated in-vivo experiments, in which the rat spinal cord was placed at different positions in a carbon SOBP [39–41]. A corresponding experiment with proton irradiation is available [42], however, their Q intervals do not overlap.

Proton therapy is rapidly expanding worldwide; clinical proton RBE data obtained from patient outcome analysis are emerging but are still sparse [2,3,10,43–45] and more clinical evidence is requested by proton therapy centers [46]. The Q model suggests that the biological effectiveness of proton and other particles, e.g. carbon ions, would be similar under the same physical and biological conditions parametrized by Q and $\alpha_x/\beta_x$, respectively. Specifically, protons and carbon ions may have similar RBE values, if the energy of carbon ions is larger by a factor of 36 (i.e., squared ratio of ion charges, Supplement figure S4). Thus, Q may support the transfer of available experience on RBE from various ions, e.g., carbon ion therapy in a simple way to fill the gap for other particles that lack sufficient RBE data. Besides proton irradiation, in future, this may also be of interest for, e.g., recently initiated helium treatments [47] or multi-ion radiotherapy [48].

The ultimate goal to predict clinical proton RBE using a Q model derived from existing clinical RBE data (mainly clinical carbon RBE), may be hampered as the clinical RBE is not only affected by biological and physical parameters, but also by institute-specific factors including dose prescription and medical decisions [49]. In this work, the experimental details (energy spectrum, secondary particles, institutional differences including biological protocols) also vary dramatically between different data points. Nevertheless, prediction of proton RBE using a simple Q model derived from the RBE of other particles was shown to be possible in-vitro by only considering the quantities energy, charge and $\alpha_x/\beta_x$. Towards future clinical application, it would be necessary to investigate whether the Q concept (RBE of different particles follows the same trend) holds in the case of clinical data. A potential clinical endpoint could be radiation-induced brain injury (manifested as image changes on follow-up magnetic resonance imaging), as clinical studies are available for patients treated with both protons [2,3,10] and carbon ions [50,51].

In conclusion, a linear model to describe variable RBE was built in a clinical proton energy range. It was based on the biological quantity $\alpha_x/\beta_x$ and the physical beam quality Q replacing the commonly used LET. A Q model built without any contribution of proton RBE data demonstrated a similar prediction power for proton $RBE_{max}$ as a widely used proton-specific model. This enables the transfer of RBE data and knowledge from heavier ions to proton therapy and facilitates the understanding of clinical proton RBE still suffering from sparse patient data. Reducing the dominant biological uncertainty in RBE modelling would, eventually, translate in considerable reduction of RBE-related uncertainty in proton therapy treatment planning.

**Figure captions**

Figure 2: Data selection workflow: Five general and one model specific filter was applied to select data from the PIDE database for the analysis of the $RBE_{max}$ (left) and $RBE_{min}$ (right). For each ion, the number of available data points is provided in brackets. All ion datasets with less than five data points were excluded from analysis. LET: linear energy transfer. RBE: relative biological effectiveness

Figure 2: The Spearman's correlation coefficient between different outputs ($RBE_{max}$, $RBE_{min}$, $\alpha_i$ and $\beta_i$) and inputs (Q, LET, $Q/(\alpha_x/\beta_x)$, $LET/(\alpha_x/\beta_x)$, $\alpha_x/\beta_x$, $\alpha_x$ and $\beta_x$) considered in this work. LET: linear energy transfer. RBE: relative biological effectiveness.

Figure 3: Linear regressions between $RBE_{max}$ and $LET/(\alpha_x/\beta_x)$ (left, LET model) or $Q/(\alpha_x/\beta_x)$ (right, Q model) for proton (A, B), helium (C, D), carbon (E, F) and neon (G, H) as well as data for all particle types (I, J). The regression lines are shown for individual ions (dashed line) and for all particles (black solid line). Values for the slope, i.e. kLET [Gy·(keV/μm)$^{-1}$] and k$_Q$ [A·MeV·Gy], and r$^2$ are provided in each subfigure. LET: linear energy transfer. RBE: relative biological effectiveness.

Figure 4: Linear regressions between $RBE_{min}$ and $LET/(\alpha_x/\beta_x)$ (left) or $Q/(\alpha_x/\beta_x)$ (right) for proton (A, B), helium (C, D) and carbon (E, F). The regression lines are shown for individual ions (solid line). Values for the slope, i.e. $S_{LET}$ [Gy·(keV/μm)$^{-1}$] and $S_Q$ [A·MeV·Gy], and r$^2$ as well as the mean and standard deviation of $RBE_{min}$ are provided in each subfigure. Note that one outlier was not included in the regression between $RBE_{min}$ and $Q/(\alpha_x/\beta_x)$, cf. D as this single point dramatically affects the fitting result. LET: linear energy transfer. RBE: relative biological effectiveness.

Figure 5: Comparison between the recorded experimental $RBE_{max}$ and the model predictions given by (A) the Q model and (B) the original Wedenberg model for all proton data considered in this work. The r$^2$ between the experimental records and the corresponding predictions as well as the identity line (y=x) as reference are provided.

Figure 6: Comparison between the recorded experimental and the predicted $\alpha_i$ based on the Q model. (A) shows all proton data considered in this work, while (B) shows a closeup restricted to values $\alpha_i \leq$ 1 Gy$^{-1}$ and effectively excluding proton data that the model failed to predict, namely, cell lines AG01522 *[7]* and HCT116 *[38]*. The r$^2$ between the experimental records and the corresponding predictions as well as the identity line (y=x) as reference are provided.

# Figures

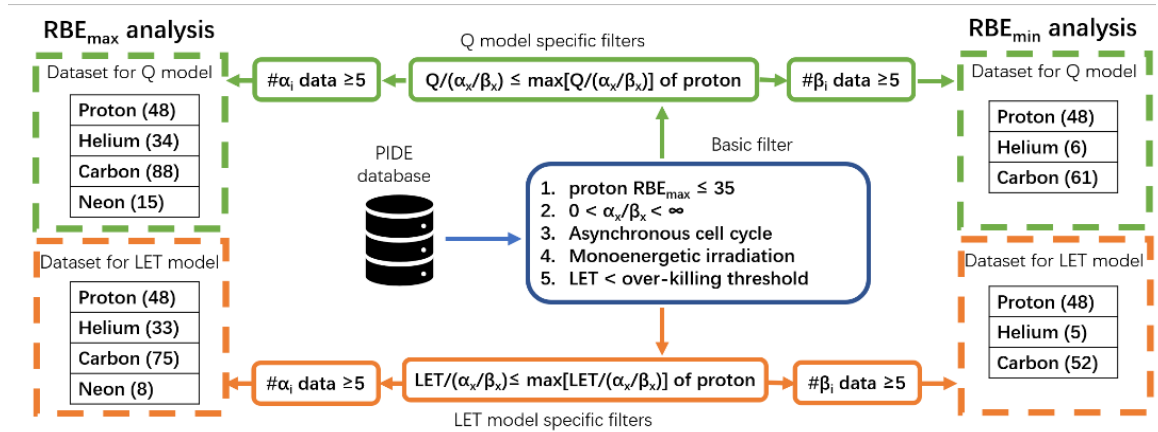

**Figure 3**: Data selection workflow: Five general and one model specific filter was applied to select data from the PIDE database for the analysis of the $RBE_{max}$ (left) and $RBE_{min}$ (right). For each ion, the number of available data points is provided in brackets. All ion datasets with less than five data points were excluded from analysis. LET: linear energy transfer. RBE: relative biological effectiveness.

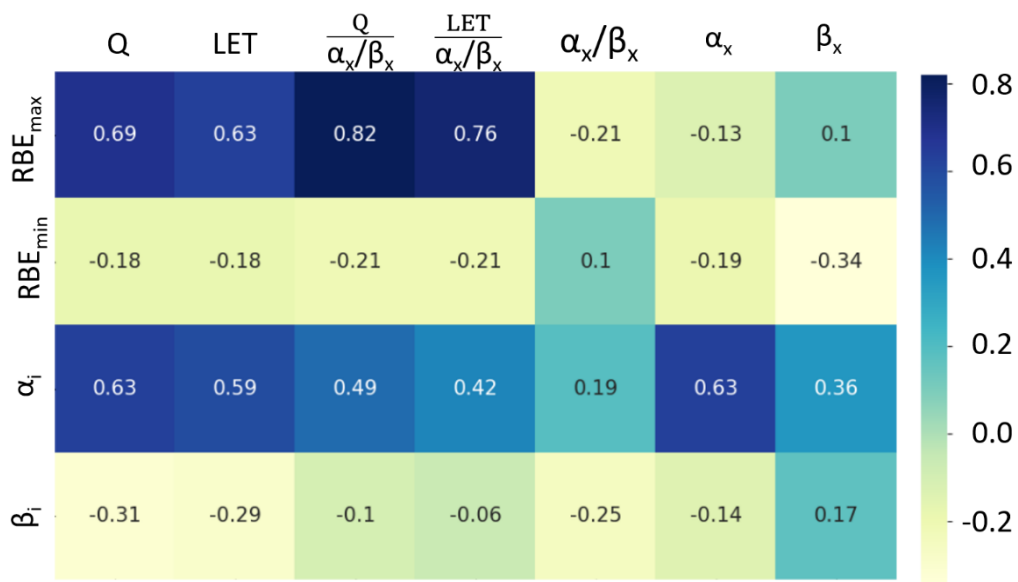

**Figure 2**: The Spearman's correlation coefficient between different outputs ($RBE_{max}$, $RBE_{min}$, $\alpha_i$ and $\beta_i$) and inputs (Q, LET, $Q/(\alpha_x/\beta_x)$, $LET/(\alpha_x/\beta_x)$, $\alpha_x/\beta_x$, $\alpha_x$ and $\beta_x$) considered in this work. LET: linear energy transfer. RBE: relative biological effectiveness.

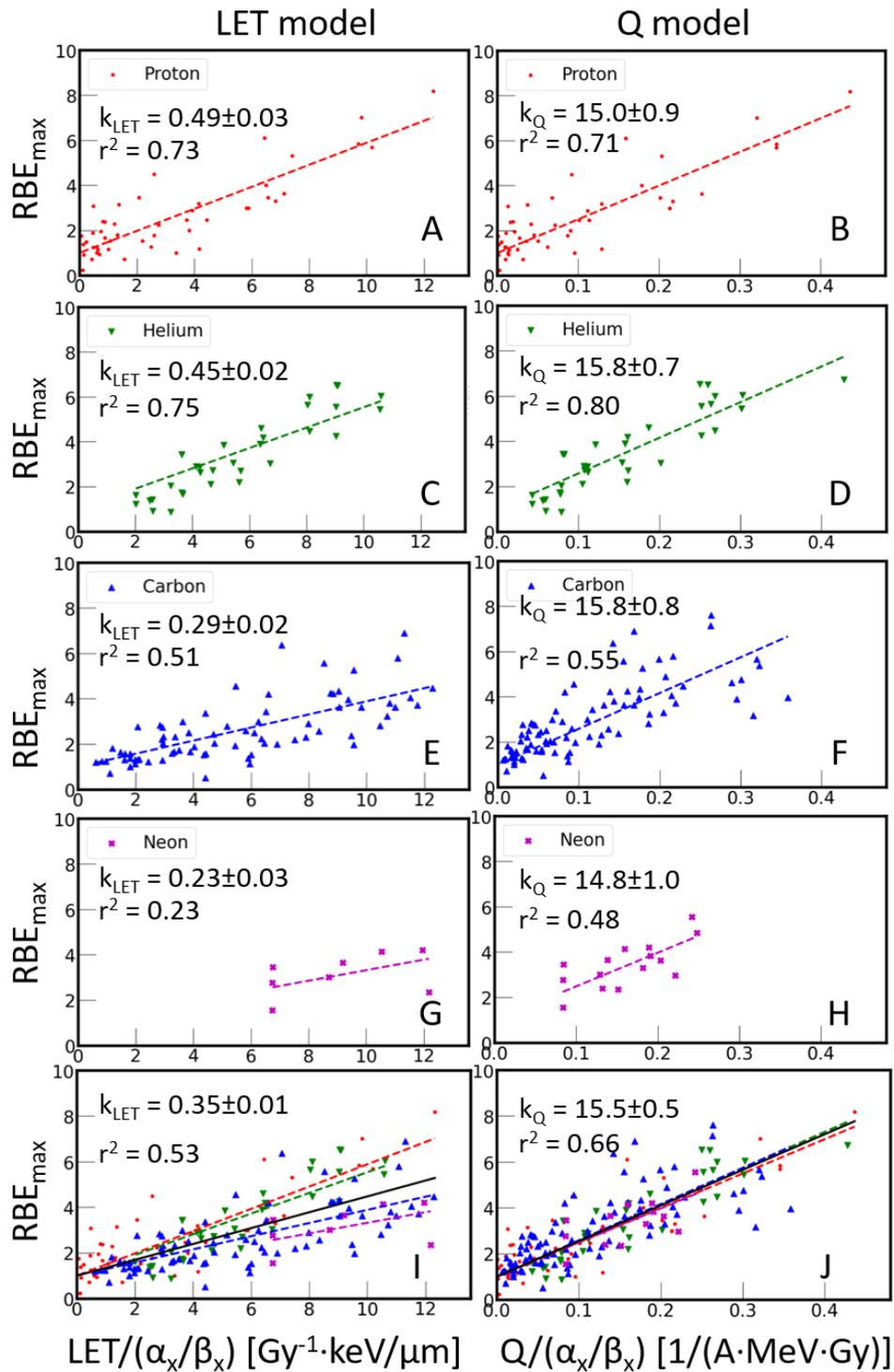

**Figure 3**: Linear regressions between $RBE_{max}$ and $LET/(\alpha_x/\beta_x)$ (left, LET model) or $Q/(\alpha_x/\beta_x)$ (right, Q model) for proton (A, B), helium (C, D), carbon (E, F) and neon (G, H) as well as data for all particle types (I, J). The regression lines are shown for individual ions (dashed line) and for all particles (black solid line). Values for the slope, i.e. $k_{LET}$ [Gy·(keV/μm)$^{-1}$] and $k_Q$ [A·MeV·Gy], and $r^2$ are provided in each subfigure. LET: linear energy transfer. RBE: relative biological effectiveness.

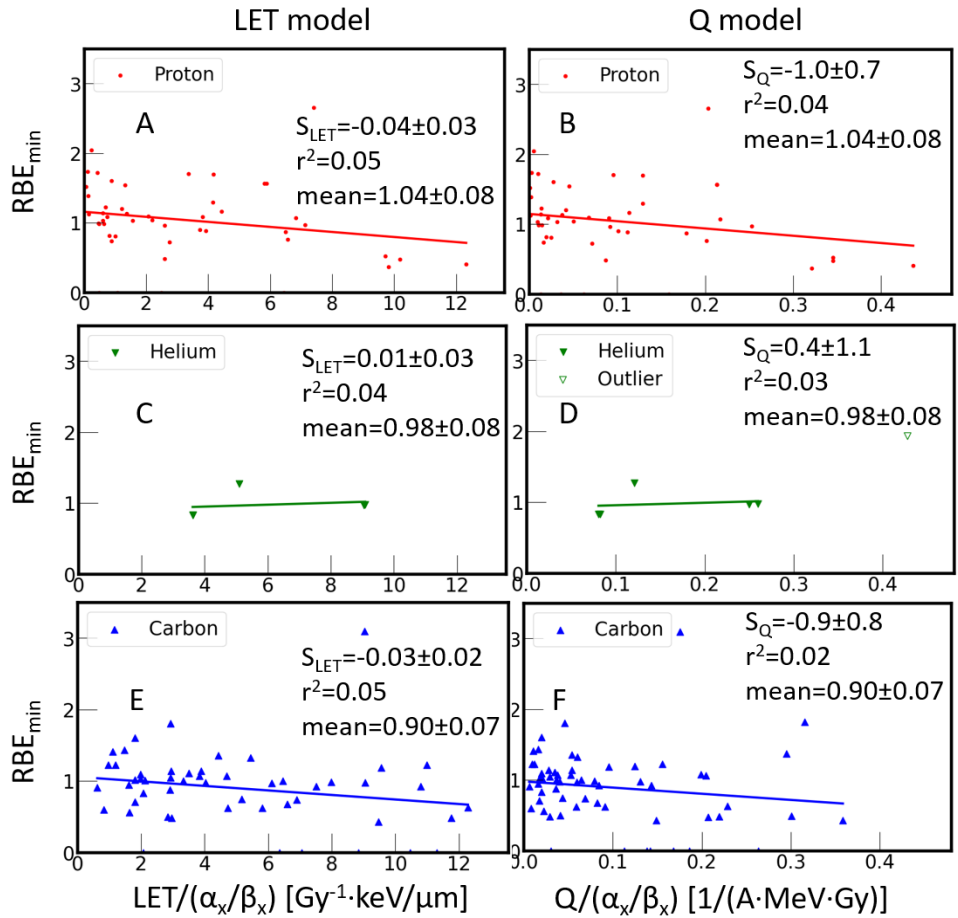

**Figure 4**: Linear regressions between RBE$_{min}$ and LET/($\alpha_x/\beta_x$) (left) or Q/($\alpha_x/\beta_x$) (right) for proton (A, B), helium (C, D) and carbon (E, F). The regression lines are shown for individual ions (solid line). Values for the slope, i.e. S$_{LET}$ [Gy·(keV/µm)$^{-1}$] and S$_Q$ [A·MeV·Gy], and r$^2$ as well as the mean and standard deviation of RBE$_{min}$ are provided in each subfigure. Note that one outlier was not included in the regression between RBE$_{min}$ and Q/($\alpha_x/\beta_x$), cf. D as this single point dramatically affects the fitting result. LET: linear energy transfer. RBE: relative biological effectiveness.

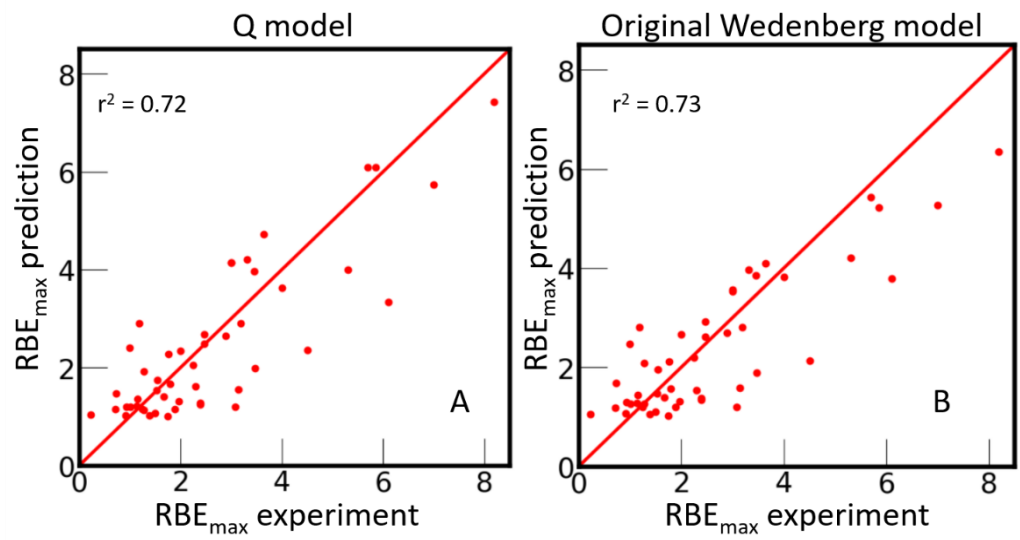

**Figure 5**: Comparison between the recorded experimental $RBE_{max}$ and the model predictions given by (A) the Q model and (B) the original Wedenberg model for all proton data considered in this work. The $r^2$ between the experimental records and the corresponding predictions as well as the identity line (y=x) as reference are provided.

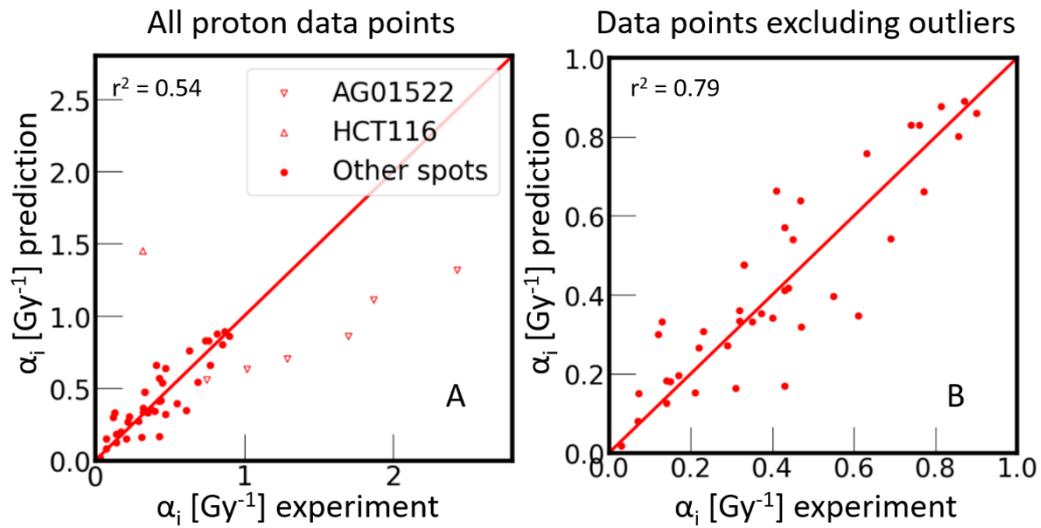

**Figure 6**: Comparison between the recorded experimental and the predicted $α_i$ based on the Q model. (A) shows all proton data considered in this work, while (B) shows a closeup restricted to values $α_i ≤ 1$ Gy$^{-1}$ and effectively excluding proton data that the model failed to predict, namely, cell lines AG01522 *[7]* and HCT116 *[38]*. The r$^2$ between the experimental records and the corresponding predictions as well as the identity line (y=x) as reference are provided.

# Supplementary Material

## RBE formulas for the Q model

Given the quantities of $\alpha_x$, $\beta_x$, $RBE_{max}$, $RBE_{min}$ and $\alpha_x/\beta_x$, the RBE can be expressed in the framework of the linear-quadratic model either as a function of the photon dose $D_x$ [S1]:

$$RBE(Q, D_x, \alpha_x/\beta_x) = \frac{RBE_{max} \cdot (\alpha_x/\beta_x) + \sqrt{(\alpha_x/\beta_x)^2 \cdot RBE_{max}^2 + 4D_x \cdot (\alpha_x/\beta_x + D_x) \cdot RBE_{min}^2}}{2(\alpha_x/\beta_x + D_x)} \quad (S1)$$

or as a function of the ion dose $D_i$ [S2]:

$$RBE(Q, D_i, \alpha_x/\beta_x) = -\frac{\alpha_x/\beta_x}{2D_i} + \frac{1}{2D_i} \cdot \sqrt{(\alpha_x/\beta_x)^2 + 4RBE_{max} \cdot \alpha_x/\beta_x \cdot D_i + 4RBE_{min} \cdot D_i^2} \quad (S2)$$

Considering the Q model proposed in this work (cf. formula (3) and (4)) to express $(\alpha_x/\beta_x) \cdot RBE_{max} = \alpha_x/\beta_x + k_Q \cdot Q$ and $RBE_{min} = 1$, the corresponding RBE formulas (S1) and (S2) result in:

$$RBE(Q, D_x, \alpha_x/\beta_x) = \frac{\alpha_x/\beta_x + k_Q \cdot Q + \sqrt{(\alpha_x/\beta_x + k_Q \cdot Q)^2 + 4D_x \cdot (\alpha_x/\beta_x + D_x)}}{2(\alpha_x/\beta_x + D_x)} \quad (S3)$$

and

$$RBE(Q, D_i, \alpha_x/\beta_x) = -\frac{\alpha_x/\beta_x}{2D_i} + \frac{1}{2D_i} \cdot \sqrt{(\alpha_x/\beta_x)^2 + 4(\alpha_x/\beta_x + k_Q \cdot Q) \cdot D_i + 4D_i^2} \quad (S4)$$

## Supplementary Tables

**Table S1**. The fitting parameters, i.e. $k_{LET}$ and $k_Q$, as well as the corresponding $r^2$ values for the LET and Q models obtained on corresponding datasets, respectively.

| Dataset | LET model | | Q model | |
|---|---|---|---|---|
| | $k_{LET}$ [Gy·(keV/μm)$^{-1}$] | $r^2$ | $k_Q$ [A·MeV·Gy] | $r^2$ |
| Proton | 0.49 ± 0.03 | 0.73 | 15.0 ± 0.9 | 0.71 |
| Helium | 0.45 ± 0.02 | 0.75 | 15.8 ± 0.7 | 0.80 |
| Carbon | 0.29 ± 0.02 | 0.51 | 15.8 ± 0.8 | 0.55 |
| Neon | 0.23 ± 0.03 | 0.23 | 14.8 ± 1.0 | 0.48 |
| Global (all ions) | 0.35 ± 0.01 | 0.53 | 15.5 ± 0.5 | 0.66 |

**Supplementary Figures**

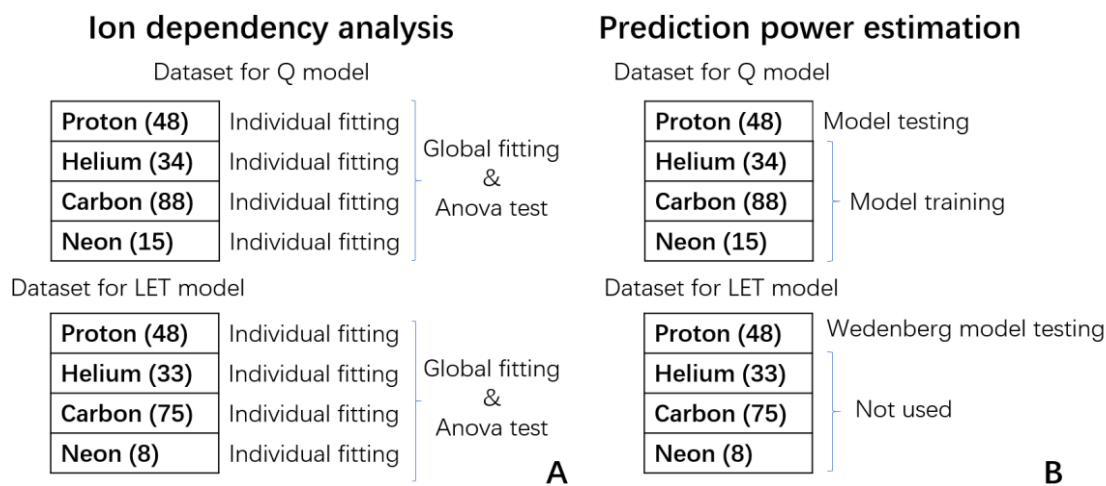

**Figure S1**: Use of datasets for different ions (number of data points) in different analysis steps. (A) Ion dependency analysis (cf. section 2.4): Linear regressions were applied both for the data of each individual ion (cf. figure 3: A-H and figure 4) and for the pooled data regardless of ion type (called global fitting cf. figure 3: black solid lines in I and J). The ANOVA tests were conducted by comparing the global fit and the data of each individual ion. (B) Prediction power estimation (cf. section 2.5): For the Q model, the data of helium, carbon and neon were used as training set while the proton data were used as test data. For the LET model, the published model parameter by Wedenberg et al. [S2] was used and the proton data were used for testing. That is, for LET, no training was performed and the data of helium, carbon and neon were not used.

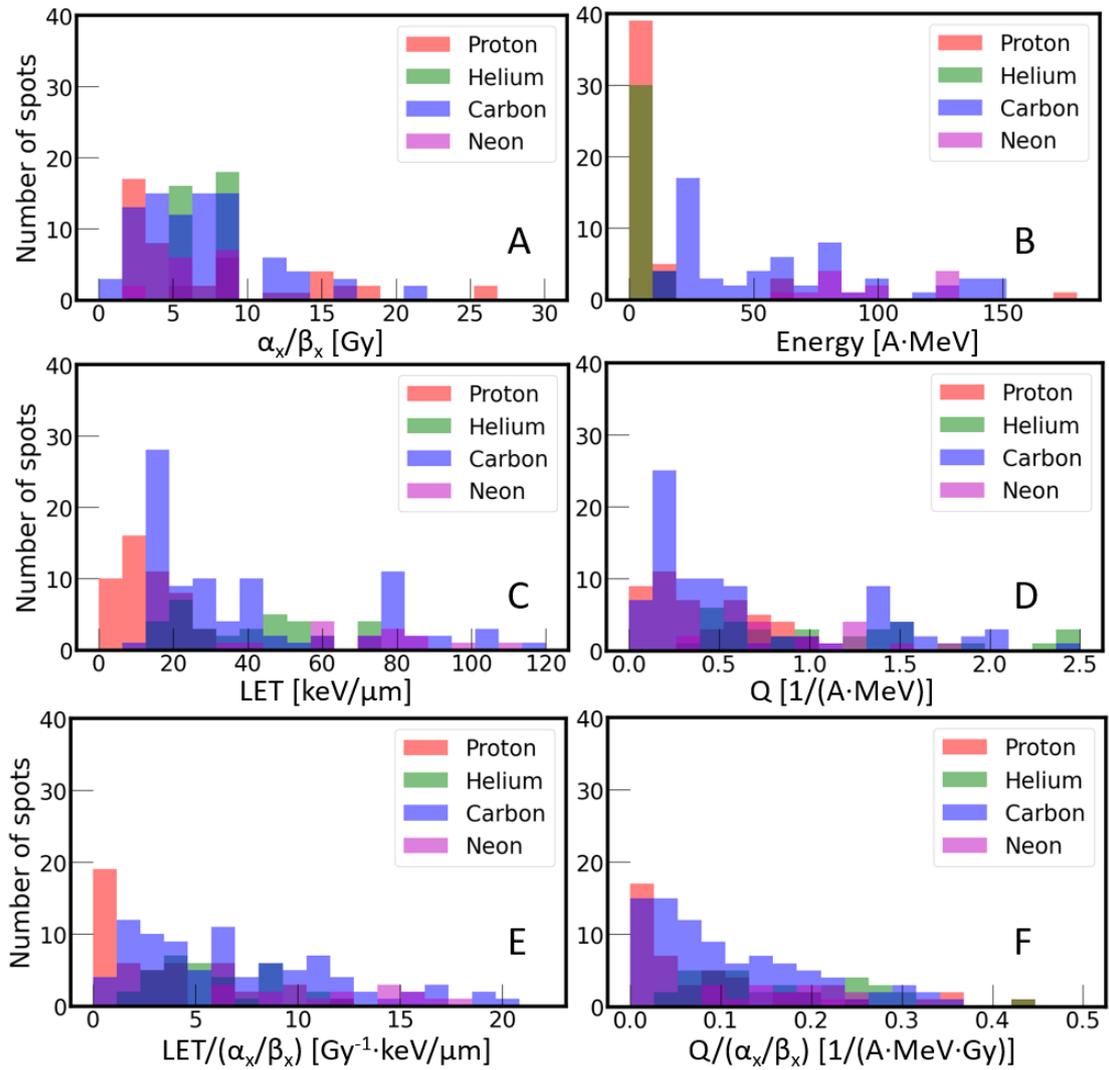

**Figure S2**: Histograms of the different quantities $\alpha_x/\beta_x$ (A), energy (B), LET (C), Q (D), LET/$(\alpha_x/\beta_x)$ (E) and Q/$(\alpha_x/\beta_x)$ (F) resolved per ion for the dataset selected for the Q model. Note, one proton irradiation data point with high $\alpha_x/\beta_x$ of 69.5 Gy was used but is not shown in subfigure A.

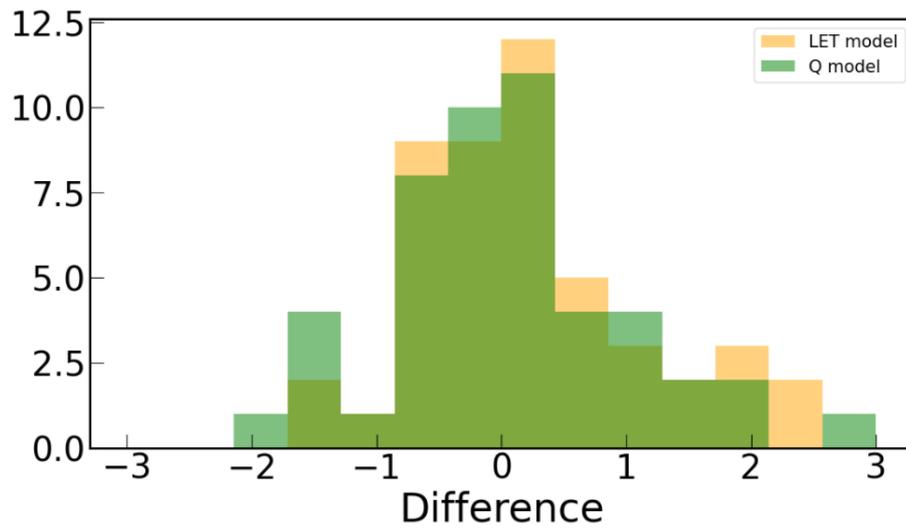

**Figure S3**: Histograms of the residuals between the experimental records and the predictions of $RBE_{max}$ based on the Q model and the LET model by Wedenberg.

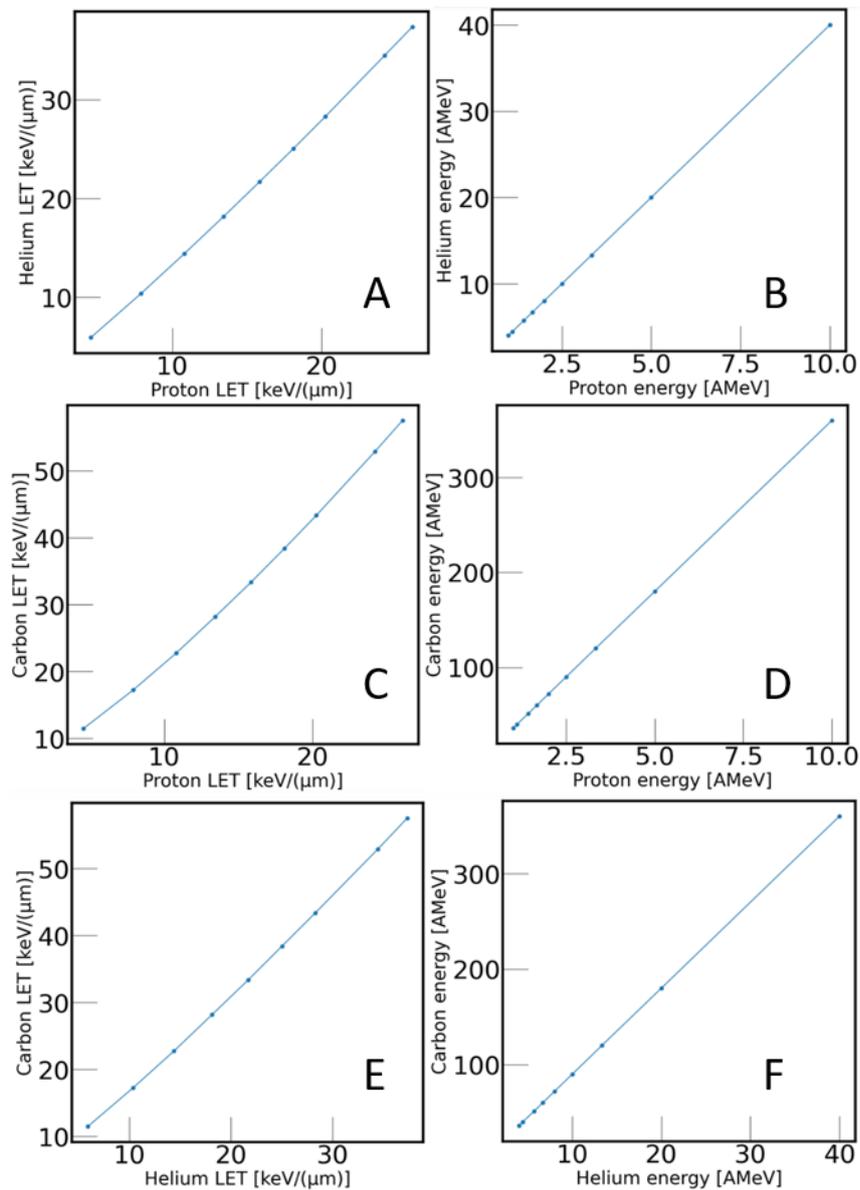

**Figure S4**: Relation of LET (left) and energy (right) values of ions: proton vs. helium (A, B), proton vs. carbon (C, D) and helium vs. carbon (E, F), resulting in the same Q values for $Q \leq 1$ $(A \cdot MeV)^{-1}$.